# Largely tunable band structures of few-layer InSe by uniaxial strain

Chaoyu Song[1,3†], Fengren Fan[1,2,3†], Ningning Xuan[4], Shenyang Huang[1,3], Guowei Zhang[1,3], Chong Wang[1,3], Zhengzong Sun[4], Hua Wu[1,2,3*], Hugen Yan[1,3*]

1. State Key Laboratory of Applied Surface Physics and Department of Physics, Fudan University, Shanghai 200433, China.

2. Key Laboratory of Computational Physical Sciences (Ministry of Education), Fudan University, Shanghai 200433, China

3. Collaborative Innovation Center of Advanced Microstructures, Nanjing 210093, China

4. Department of Chemistry and Shanghai Key Laboratory of Molecular Catalysis and Innovative Materials, Fudan University, Shanghai 200433, China

† These authors contributed equally

* Emails: hgyan@fudan.edu.cn (H. Y.), wuh@fudan.edu.cn (H. W)

# Abstract

Due to the strong quantum confinement effect, few-layer γ-InSe exhibits a layer-dependent bandgap, spanning the visible and near infrared regions, and thus recently draws tremendous attention. As a two-dimensional material, the mechanical flexibility provides an additional tuning knob for the electronic structure. Here, for the first time, we engineer the band structures of few-layer and bulk-like InSe by uniaxial tensile strain, and observe salient shift of photoluminescence (PL) peaks. The shift rate of the optical gap is approximately 90-100 meV per 1% strain for 4- to 8-layer samples, which is much larger than that for the widely studied $MoS_2$ monolayer. Density functional calculations well reproduce the observed layer-dependent bandgaps and the strain effect, and reveal that the shift rate decreases with increasing layer number for few-layer InSe. Our study demonstrates that InSe is a very versatile 2D electronic and optoelectronic material, which is suitable for tunable light emitters, photo-detectors and other optoelectronic devices.

Key words:  Uniaxial strain, InSe, Photoluminesence, layer-dependent, DFT calculation

2-dimensional (2D) van der Waals atomic crystals typically exhibit very different properties from their bulk counterparts. Since the isolation of graphene[1] in 2004, the research field of 2D materials has exponentially grown. In the post-graphene era, much attention is paid to the exploration of new 2D materials, such as transition metal dichalcogenides[2] (TMDCs), hexagonal boron nitride[3], silicene[4], stanene[5] and black phosphorus[6, 7]. Recently, another layered metal chalcogenide semiconductor, γ-phase indium selenide (γ-InSe), was exfoliated to atomically thin layers and gathers interest of the scientific community[8]. The electronic and optical properties of Bulk InSe[9] were first studied more than half a century ago. It possesses a direct bandgap of 1.25eV and shows anisotropic electronic properties originated from the layered structure. When decreasing the thickness from bulk to monolayer, the bandgap increases over 0.5eV due to quantum confinement in the out-of-plane direction, which has been confirmed by theoretical calculations[10, 11] and optical spectroscopy[8, 12-14]. Moreover, the carrier mobility exceeds $10^3$ cm$^2$ V$^{-1}$ s$^{-1}$ at room temperature[15] and quantum Hall effect has been demonstrated in high quality few-layer electronic devices[14]. Owing to its high carrier mobility, layer-tunable bandgap and ambient stability, few-layer InSe is a competitive choice for the applications in electronics and optoelectronics[16, 17].

Recently, much effort has been devoted to tuning the electronic and optical properties of InSe through various schemes, such as magnetic field[18], controllable oxidation[19], texturing[13] and high pressure[20]. In addition to these means, mechanical strain is a simple yet effective and repeatable way to continuously tune the band structure of 2D materials. The strain effect on vibrational and electronic properties of some typical 2D materials like graphene[21, 22], TMDCs[23-27] and black phosphorus[7, 28] has been extensively studied. However, up to date, systematic experimental study for strained few-layer InSe is still lacking. Moreover, previous studies for strained 2D materials mainly focus on the single layer and/or bilayer cases and pay little attention to the layer-dependent strain effect.

In this work, for the first time, we investigate the influence of uniaxial tensile strain on the electronic and optical properties of few - layer InSe with sample thickness ranging from 4 to 8 layers. In combination with DFT calculations, we reveal that the strain effect for 1- to 8-layer samples is quantitatively different, with the strongest effect for the monolayer case. This can be understood based on the strain - induced change of the inter-layer interactions. Unlike many other 2D materials, the inter-layer interaction in InSe is strong and hence causes the large variation of the bandgaps from monolayer to the bulk. Therefore, any change of the interlayer coupling can show effect on the bandgaps of few-layer InSe. Our PL measurements show that the optical bandgap of few-layer InSe decreases by approximately 90-100 meV per 1% uniaxial tensile strain. Surprisingly, by taking advantage of the small Young's modulus[29], bulk-like flakes can be tuned as well by the same experimental technique, with a strain-induced shift comparable to or even slightly greater than that for few-layer InSe, which is also consistent with DFT calculations. The strain effect on another transition (B transition shown in Fig. 1d) of the bulk-like InSe is also studied, exhibiting a smaller shift than that of the bandgap transition. With such large strain tunability, in conjunction with the strong layer-dependent band structure, few-layer InSe bandgap can continuously cover from part of the visible to the near IR wavelength range, possibly from 0.6 to 1.1microns.

## Results

**Layer-dependent band structure and optical properties**

The crystal structure of γ-InSe is depicted in Fig. 1a. In each layer, the hexagonal lattice consists of four atomic planes which are arranged in the sequence of Se-In-In-Se with covalent bonds connecting the atoms. Individual layers are stacked together through van der Waals interaction. Few-layer InSe was first exfoliated onto polydimethylsiloxane (PDMS) from the bulk crystal (2D Semiconductors Inc.) by scotch tape method[1] and then transferred to a flexible polypropylene (PP) substrate with thickness of 0.3mm. The layer thickness of InSe was determined by optical contrast[30] (see Fig. S1) and further verified by atomic force microscope (AFM) and PL spectroscopy. Raman spectroscopy was performed for several typical samples and all characteristic Raman modes were observed (See Fig. S2).

PL spectroscopy has been widely employed to study exciton emission in semiconductors. Those excitons are mainly band-edge excitons associated with direct bandgaps. The emission energy, usually termed as optical bandgap, is typically smaller than the electronic single particle bandgap (transport bandgap) with the difference as the exciton binding energy. Nevertheless, PL spectroscopy still provides us a good scheme to probe the band structures. This is particularly true for few-layer InSe, which possesses a strongly layer-dependent direct or quasi-direct bandgap[18]. As a result, the PL spectrum varies with layer thickness, as detailed in Fig. 1b. The bandgap transition of monolayer InSe is forbidden by selection rules originated from mirror-plane symmetry[11, 14]. The PL peak energy of bilayer InSe is around 1.9eV according to previous studies[13, 14]. Unfortunately, in our measurements it was too weak to retrieve from the overwhelming substrate signals. The PL spectra and peak positions of 3- to 8-layer on PP substrate and bulk InSe are illustrated in Fig. 1b and 1c, with a 532nm laser as the excitation light source. For clarity, the intensities of the PL spectra are normalized to show the same peak height in Fig. 1b. As the thickness decreases from bulk to 3 layers, the PL intensity is in fact reduced dramatically and the emission peak blueshifts from 1.25eV to 1.69eV, which is consistent with previous reports[8, 12-14]. All experiments were conducted at room temperature and under ambient conditions. We checked the PL spectra during an extended period and found that the sample quality didn't change for at least several weeks under ambient condition, indicating good air stability. Nevertheless, all data reported in this paper were acquired within half a day after exfoliation.

To further confirm the thickness dependent electronic properties, we performed DFT calculations (see Methods and S5). The detailed band structures of 1-, 3-, 5-layer and bulk InSe are shown in Fig. S5 for comparison. The calculated thickness dependent bandgaps are summarized in Fig. 1c and they overall agree well with the experiments, showing a decreasing bandgap as the thickness increases. Of course, care must be taken that the PL peaks are associated with optical bandgaps, and the DFT gives us single particle bandgaps. Therefore, comparing the absolute values is not informative, and the trend of the bandgap is more significant, which indicates a strong quantum confinement effect. We show in Fig. 1d the orbital resolved band structure of monolayer InSe. We find that the topmost valence band consists of Se $4p_z$ state and it is followed by the lower Se $4p_{x,y}$ bands, and the bottom conduction band originates from In 5s state. In Fig. 1d, we also indicate the transition A (bandgap transition) and transition B, with transition B referring to the transition from the bottom of the conduction band to the top of the Se $4p_{x,y}$ valence. The A and B transitions in other thickness samples follow the same convention.

**Strain engineering on few-layer InSe**

One of the most attractive attributes of 2D materials over their 3D counterparts is their mechanical stretchability, with a breaking strain typically above 10%[31]. Compared to other 2D materials, such as graphene and $MoS_2$, the Young's modulus[29] of thin InSe layers is much smaller, which makes strain engineering even easier. Strain changes the bond lengths of the material and hence the hopping integrals between constituent atoms. Therefore the electronic bands change accordingly. There are multiple routes to apply strain on 2D materials, with both uniaxial and biaxial strains[31] possible. In our work, controllable uniaxial strain ($\varepsilon$) was applied by two-point bending method as illustrated in Fig. 2a. The strain $\varepsilon$ in the stretch direction was determined based on the deflection of the PP substrate[25] as detailed in Fig. S3, while the strain in the other direction was neglected. The band structure and optical properties of few-layer InSe are prominently engineered under strain according to our PL measurements. Here, we only focus on transition A for few-layer InSe with layer number below 10, as the PL intensity of transition B is much weaker and poses challenge for measurements. The strain is reversible and multiple rounds of straining/releasing process typically give the same results, indicating effective strain transfer from the substrate to few-layer InSe. Fig. 2a and 2b shows the evolution of PL spectrum of 4- and 5- layer InSe while strain is applied from zero to 1.15%. The peak energy, intensity and width of the PL peak are determined by fitting the spectrum with a Lorentzian lineshape. There is no systematic change of peak intensity and width (Fig. S4). However, a prominent redshift of PL peak energies under strain can be observed. For 4-layer InSe, the PL peak shifts from 1.55 eV to 1.44 eV under 1.15% tensile strain, while for 5-layer InSe, the PL peak shifts from 1.48 eV to 1.37 eV. The PL peak positions decrease linearly with tensile strain, as shown in Fig. 2c for 4- to 8-layer InSe. We measured 3-6 samples for each thickness InSe, and the fastest shift rate is summarized in Fig. 4b. The shift rate of 4- and 5- layer InSe is about 100meV/% and 99meV/%. For 4- to 8-layer samples, the shift rate reduces slightly as the thickness increases. Compared to the widely studied MoS$_2$ monolayer (shift rate around 50-60meV/%)[31], the shift rate for few-layer InSe is much larger, indicating more efficient bandgap engineering for potential tunable optoelectronic devices. It should be noted, however, we directly probe the optical bandgap, which depends on the single particle bandgap and exciton binding energy. Under strain, the exciton binding energy typically exhibits little change[32] and hence we mainly attribute the observed PL peak shift to the modification of the single particle bandgap. This is particularly true for multilayer samples, whose exciton binding energy is much smaller compared to the monolayer and the strain effect on the exciton binding energy can be even more safely neglected.

**Strain engineering on bulk-like InSe**

The Young's modulus of InSe has been predicted to be much smaller than many other 2D materials, such as graphene and MoS$_2$. The modulus is only 10% of that for graphene[29]. As a result, it's easier to exert strain on InSe flakes sitting on flexible substrates and strain can be transferred to much thicker samples through substrate deformation. Indeed, this is the case in our study. We were able to engineer the band structure of bulk-like InSe flakes, whose thickness exceeds 50 layers and the electronic structure is the same as the bulk, but the mechanical flexibility is still comparable to the few-layer. This provides us a unique opportunity to engineer the electronic structure of a bulk material through uniaxial strain, using the simple substrate deformation method. In addition, the luminescence intensity of the bulk-like InSe is much stronger, therefore, besides the PL of transition A, we can clearly observe the PL evolution of transition B under strain. The PL spectra of transition A and B were recorded by an InGaAs detector and Si CCD (charge coupled device) respectively (see Methods).

As depicted in Fig. 3a, the PL peak of transition A of the bulk-like InSe is 1.25eV without strain, and shifts to 1.11eV when 1.15% tensile strain is applied. The excitation laser wavelength is 473nm in the measurements of transition B and the PL intensity is at least three orders of magnitude lower than that of transition A. Fig. 3b shows that the peak shifts from 2.43eV to 2.38eV. The spectral linewidth and intensity remains almost the same in the straining process for both transitions. The shift rate of transition A is 118 meV/% and that of transition B is 43meV/%, only one third of the rate for transition A. Empirically, the electronic properties of bulk-like materials should be difficult to tune by such stretching method. However, our results show that the strain effect on thick InSe films is comparable to or even slightly larger than the effect on few-layer InSe.

**DFT calculations of the strain effect**

The mechanism underlying the strain effect on InSe is investigated by DFT calculations. We simulated the strain effect in DFT calculations by expanding the lattice along the a-axis up to 2% (in a step of 0.5%). For clarity, we show the monolayer band structures here. The calculated band structures of monolayer InSe without strain (red lines) and with 2% strain (blue lines) are compared in Fig. 4a.

Upon a tensile strain, the lengths of the intralayer In-Se bonds increase , though In-In bonds have negligible change, given the bond direction perpendicular to the strain, as can be seen in Fig. 1a. The weakened bond interactions raise the bonding Se $4p_z$ valence band and lower the anti-bonding In 5s conduction band, thus giving rise to the observed decrease of the bandgap.

Since the precise experimental results of 1- to 3- layer InSe are lacking due to technical difficulties, the DFT calculations are good complements to help us understand the strain effect on all layers of InSe and show the layer dependent effect. The evolution of the calculated bandgaps under strain (independent of the scissor correction) for 1- to 8-layer and bulk InSe is summarized in Fig. 4b. For example, the scissor-corrected bandgap of monolayer InSe decreases from 2.36 eV to 2.20 eV with 2% tensile strain, while for bilayer InSe, it shifts from 2.07 eV to 1.92 eV. The bandgaps decrease almost linearly as a function of strain. The fitted shift rates for all 9 cases (blue squares) are shown in Fig. 4c. For few-layer InSe, the shift rates decrease gradually as the number of layers increases. The calculated shift rate is approximately 80 meV/% for monolayer InSe and reduces to 60 meV/% for 8-layer InSe. For bulk InSe, the shift rate is 66 meV/%, a little larger than 8-layer case. For comparison, Fig. 4c also shows the experimentally extracted shift rates for 4- to 8-layer and bulk-like samples. Though with sizable uncertainties, the experimental results are consistent with the predicted layer-dependent trend by DFT.

## Discussions

The thickness-dependence of the shift rates for few-layer InSe possibly originates from the modification of interlayer interactions under strain. The conduction bands and valence bands of few-layer InSe split into multiple subbands due to quantum confinement associated with interlayer interactions[8], which can be seen in Fig. S5. This accounts for the decreasing bandgaps with increasing film thickness (Fig. 1c). If we assume that strain doesn't affect the interlayer coupling, the splitting of valence (conduction) band in few-layer InSe due to interlayer coupling remains the same under strain, therefore the bandgap shift rate will be the same as that in the monolayer case. A reduced shift rate with increasing layer number indicates that the subband splitting decreases, and a weakened interlayer coupling can be inferred due to the in-plane tensile strain.

In addition to the bandgap transitions, the physical origin underlying the smaller shift rate of transition B in bulk-like InSe can be qualitatively explained as well. Since transitions A and B share the same conduction band edge as their emission initial state, we only have to compare the valence band final states under strain. As indicated by the DFT calculated band structures in Fig. 4a, the Se $4p_z$ valence band moves up apparently under a tensile strain, while the top-most Se $4p_{x,y}$ band energy undergoes almost no change around Γ point, though the degeneracy breaks down. This results in a slower shift rate for the $4p_{x,y}$ type transition (transition B). Though Fig. 4a is for monolayer case, bulk-like InSe behaves qualitatively the same from our calculations.

In conclusion, as a promising optoelectronic material, the band structures of few-layer InSe can be largely tuned by the uniaxial tensile strain. For 4-8 -layer InSe, the shift rate of the PL transition A is about 90-100 meV/% and decreases slowly as thickness increases. The DFT calculated shift rates are comparable with the experimental results. The modification of interlayer coupling due to strain is a possible origin of the layer-dependent shift rate. Our study shows that few-layer and bulk-like InSe flakes are highly tunable optoelectronic materials, which covers a wide spectral range from part of the visible to the near IR.

# Methods

**PL spectroscopy** The PL measurement of few-layer InSe (3-8 layer) was taken by a Horiba iHR550 spectrometer with a silicon CCD detector array. A 532 nm laser (spot size ~1μm, laser power of 20-100μW) was used as the excitation source. For thick InSe flakes (over 50 layer), we used an Andor SR500i spectrometer equipped with an InGaAs detector and 532nm laser (spot size ~1μm, laser power of 400μW) to measure the PL of transition A, and Horiba HR800 Raman system with 473nm laser (spot size ~1μm, laser power of 500μW) to measure the PL of transition B.

**DFT Calculation Methods** The DFT calculations are performed using the Vienna *ab initio* simulation package (VASP)[33, 34] with the Perdew-Burke-Ernzerhof (PBE) GGA[35] functional, and the Van der Waals interaction is included using the optB88 vdW density functional[36, 37]. The energy cutoff is set to be 500 eV. For few-layer samples, a 18×18×1 Monkhorst-Pack grid of k-points is used; but for bulk calculations, a 18×18×6 grid instead. Test calculations using the Heyd-Scuseria-Ernzerhof hybrid functional (HSE) are also performed, see Fig. S6 and Note S6. It turns out that the GGA plus optB88 functional is economic and sufficiently accurate here for studying the band structures and the shift rates.

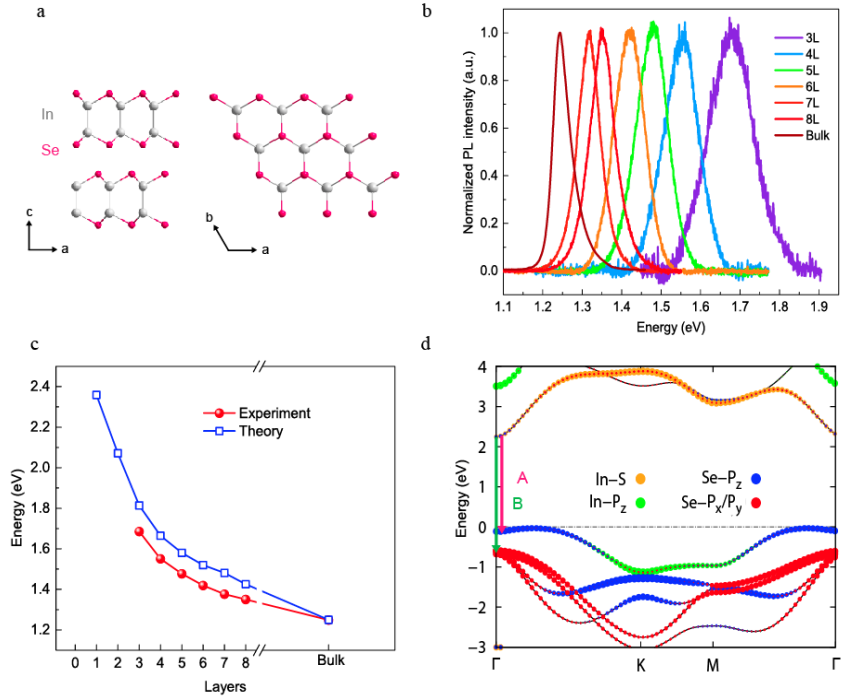

**Figure 1.** (a) Crystal structure of bilayer InSe viewed from b-direction (left) and monolayer InSe viewed from c-direction (right), the red and grey circles denotes Selenium and Indium atoms, respectively. (b) Normalized PL spectra of Bulk and 3- to 8-layer InSe. (c) The PL peak energy and DFT calculated bandgap (scissor corrected) of different thickness InSe. (d) Orbital resolved band structure of monolayer InSe, The orange, green, blue, and red dots denotes the In-S, In-$p_z$, Se-$p_z$, and Se-$p_{x,y}$, respectively. The size of the dot gives the weight of corresponding orbital components. A and B transitions are indicated as pink and green arrows respectively.

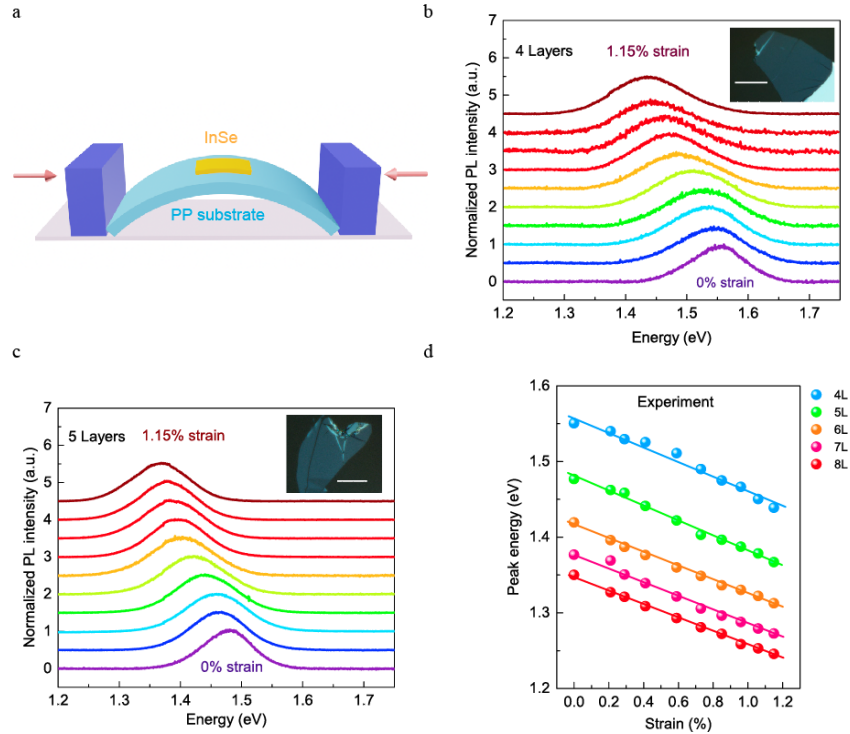

**Figure 2.** (a) Schematic illustration of the two-point bending apparatus. (b) and (c) Normalized PL spectra of transition A of 4-and 5-layer InSe, Inset: Sample optical images with the scale bar of 20 μm. (d) The PL peak energies of 4-,5-,6-,7- and 8-layer InSe as a function of uniaxial strain. Solid lines are linear fits.

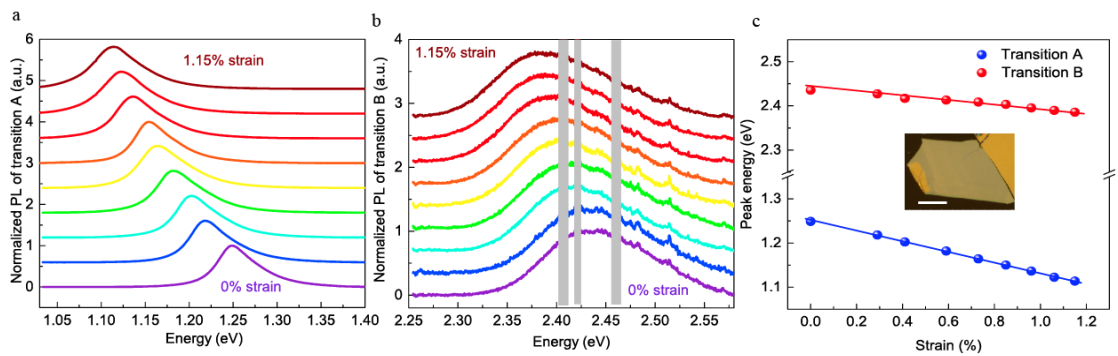

**Figure 3.** (a) Normalized PL spectra of transition A of bulk-like InSe under strain. (b) Normalized PL spectra of transition B of thick InSe under strain, the grey areas corresponds to PL or Raman signals from the substrate, which we intentionally removed. (c) The comparison of peak positions of transition A and B under different

strains. Solid lines are linear fits. Inset: a typical sample image with the scale bar of 20 μm.

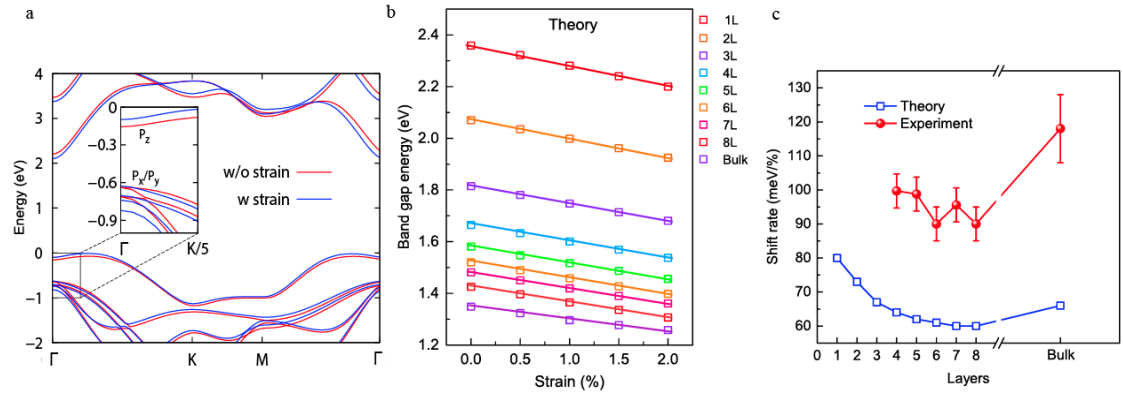

**Figure 4.** (a) DFT calculated Band structures of monolayer InSe without (red lines) and with (blue lines) 2% uniaxial strain, the inset is a zoom-in of the valence bands. (b) The energies of DFT calculated bandgaps under various strains for 1- to 8-layer and bulk InSe. The solid lines are linear fits. (c) The strain shift rates of few-layer and bulk-like InSe from PL measurements and DFT calculations.